# Strain-induced exciton transfer among quantum emitters in two-dimensional materials

*Giuseppe Ronco, Abel Martínez-Suárez, Davide Tedeschi\*, Matteo Savaresi, Aurelio Hierro-Rodríguez, Stephen McVitie, Sandra Stroj, Johannes Aberl, Mortiz Brehm, Victor M. García-Suárez, Michele B. Rota, Pablo Alonso-González, Javier Martín-Sánchez\*, and Rinaldo Trotta\**

G. Ronco, D. Tedeschi, M. Savaresi, M. B. Rota, R. Trotta

Department of Physics, Sapienza University of Rome, Piazzale A. Moro 5, 00185 Rome, Italy

E-mail: rinaldo.trotta@uniroma1.it

D. Tedeschi

ENEA, C.R. Casaccia, via Anguillarese 301, 00123 Roma, Italy

E-mail: davide.tedeschi@enea.it

A. Martínez-Suárez, A. Hierro-Rodríguez, Victor M. García-Suárez, P. Alonso-Gonzáles, J. Martín-Sánchez

Department of Physics, University of Oviedo, C/ Federico García Lorca nº 18,

33007, Oviedo, Spain

Nanomaterials and Nanotechnology Research Center (CINN-CSIC), Av. la Vega, 4, 6,

33940, El Entrego, Spain

E-mail: javiermartin@uniovi.es

A. Hierro-Rodríguez, Stephen MC Vitie

SUPA, School of Physics and Astronomy, University of Glasgow, G12 8QQ, Glasgow, U.K.

S. Stroj

Forschungszentrum Mikrotechnik, FH Vorarlberg, Hochschulstr. 1, A-6850, Dornbirn,

Austria

J. Aberl, M. Brehm

Institute of Semiconductor and Solid State Physics, Johannes Kepler University Linz,

Altenbergerstraße 69, 4040 Linz, Austria

Keywords: Transition Metal Dichalcogenides, single-photon sources, strain tuning, exciton diffusion

The discovery of quantum emitters (QEs) in two-dimensional materials (2D) has triggered a surge of research to assess their suitability for quantum photonics. While their microscopic origin is still the subject of intense studies, position-controlled QEs are routinely fabricated using static strain gradients, which are used to drive excitons towards localized regions of the crystal where quantum light emission takes place. However, the use of strain in a dynamic fashion to control the brightness of single-photon sources in 2D materials has not been explored so far. In this work, we address this challenge by introducing a novel hybrid semiconductor-piezoelectric device in which WSe$_2$ monolayers are integrated onto piezoelectric pillars that provide both static and dynamic strains. The static strains are first used to induce the formation of QEs, whose emission shows photon anti-bunching. Their energy and brightness are then controlled via the application of voltages to the piezoelectric pillars. Numerical simulations combined with drift-diffusion equations show that these effects are due to a strain-induced modification of the confining-potential landscape, which in turn leads to a net redistribution of excitons among the different QEs. Our work provides a method to dynamically control the brightness of single photon sources based on 2D materials

## 1. Introduction

The first seminal works reporting single photon emission in transition metal dichalcogenides (TMDs) [1-5] have stimulated an explosion of research activities aimed at investigating the possibility to use quantum emitters (QEs) in TMDs for ultra-compact quantum photonics. [6,7] Compared to other established solid-state-based quantum-light sources, such as semiconductor quantum dots, QEs in TMDs have the clear advantage of being relatively simple and cheap to fabricate, and their spatial position across the substrate can be controlled with high precision.[8-12] In addition, well-established processing techniques developed for conventional semiconductors can also be exported to TMDs so as to achieve, for example, coupling of QEs with nanophotonic cavities or tuning of QE emission properties via external perturbations.[13-19] Yet, QEs in TMDs still need to prove their real potential for quantum photonics: Even though single photon emission in TMDs can be routinely observed, the low photon indistinguishability[20] and coherence time[21] are limiting factors. Even the generation of entangled photons – a possibility suggested by recent works but hampered by the presence of a sizeable exciton fine structure splitting – has still to be demonstrated.[22,23] Therefore, it is quite clear that additional research activities aimed at understanding the origin and the fundamental properties of QEs in TMDs as well as at developing novel source-engineering methods are paramount.

The formation of QEs in TMDs has been observed in WSe2, WS2, MoS2 and MoTe2, using a variety of methods.[8,10,12,24,25] Most of them make use of static strain-gradients that switch on exciton funnelling towards localized potential wells where single photon emission takes place. Whether strain alone is sufficient to create these potential wells, or it needs the aid of defects to enable the formation of localized intervalley bound states is still a question of theoretical debate.[26,27,28] From the experimental side, on the other hand, strain gradients that allow for the formation of QEs are usually obtained upon transferring thin TMD crystals fabricated via mechanical exfoliation on textured substrates featuring nanopillars, metal nanostructures, nano-indentations and nanobubbles[8,29-32] to mention a few. Recent experiments using an atomic force microscopy (AFM) tip have also shown that it is possible to attain tight control over the strain profile, and the deterministic writing of QEs in TMDs has become reality.[29] However, in all these schemes, the strain configuration is usually frozen. This leads to QEs whose emission properties are fixed by the local degree of bending of the monolayer (ML), i.e., by the local strain configuration that has enabled their formation. Moreover, different QEs feature dissimilar emission properties (including energy, intensity, and polarization) due to different local strain configurations at the QE location. This is clearly not ideal for several advanced quantum photonic applications which instead require photonic states with the same properties. And previous attempts to strain tuning of TMDs have indeed demonstrated the possibility to attain dynamic control over the exciton emission energy and polarization angle.[13,33,34] However, considering the key role of strain, a natural question arises: can strain be used to dynamically control the brightness of QEs in TMDs?

In this work, we demonstrate experimentally and theoretically an important aspect that has been overlooked in previous works due to the lack of suitable technological platforms: the possibility of using dynamic strain fields to control the distribution of excitons among individual QEs. This is achieved via a novel device featuring piezoelectric pillars onto which a WSe2 ML is transferred. Specifically, we show that the application of a voltage to the device allows us to tune not only the energy but also the brightness of QEs in TMDs. Numerical simulations combined with drift-diffusion equations show that the experimental findings can be explained by a strain-induced reconfigurable potential landscape with localized potential wells, whose confining energy can be modified to change the QEs being populated by the photo-generated carriers.

## 2. Results

### 2.1. Site-controlled and energy-tunable quantum emitters on piezoelectric pillars

The working principle of the hybrid 2D-semiconductor-piezoelectric device used in this work is illustrated in **Figure 1a**: it consists of a gold-coated (001)-[Pb(Mg1/3Nb2/3)O3]0.72-[PbTiO3]0.28 (PMN-PT) piezoelectric plate with a WSe$_2$ ML attached to it by Van der Waals forces (see methods). The top side of the piezoelectric plate is electrically grounded, whereas a voltage is applied on the bottom side: an induced vertical electric field ($F_p$) along the poling direction of the PMN-PT produces an out-of-plane (strain along the z-direction) and an in-plane (strain in the x-y direction) deformation of the plate, which is then transferred to the attached ML. The amplitude and polarity of the applied voltage control the magnitude and the compressive/tensile nature of the strain field, respectively.[14,33] To induce the static deformations of the ML at predefined positions, i.e., the formation of site-controlled QEs (8), an array of piezoelectric pillars is fabricated by focused ion beam (FIB) (see methods). The diameter (height) of the pillars is about 1.4 μm (115 nm), and they are arranged across the substrate in a square pattern.

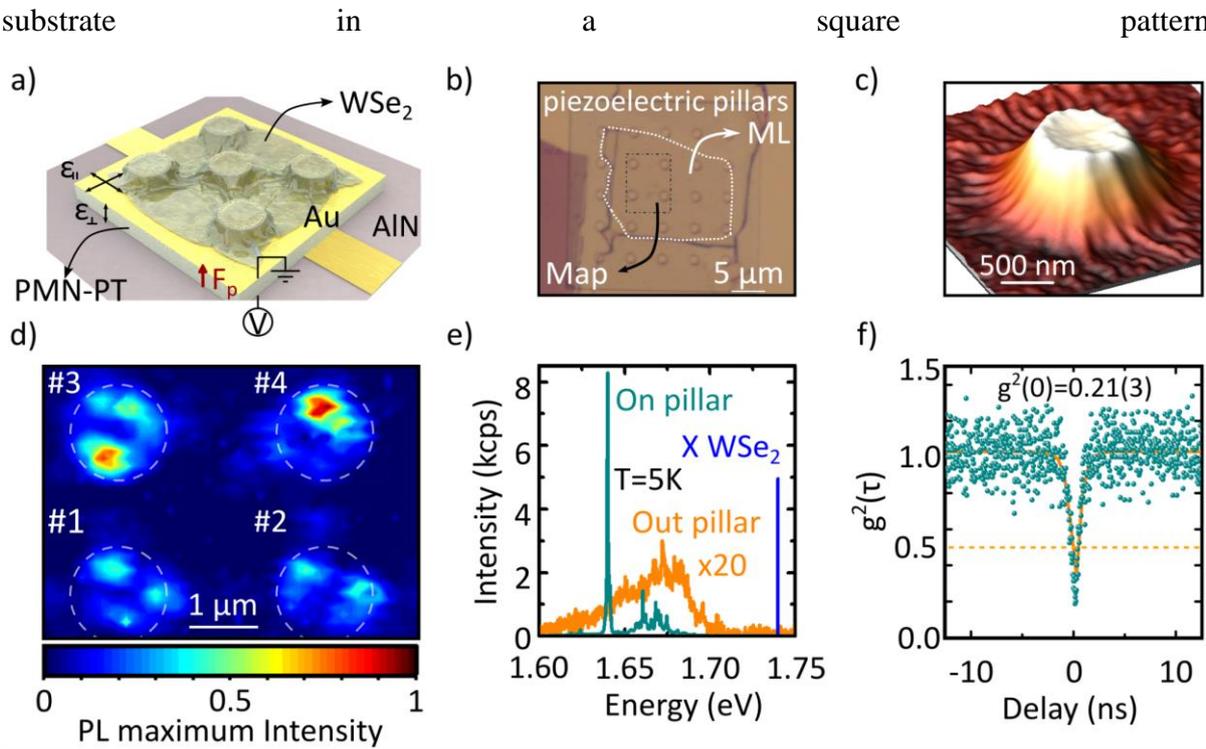

**Figure 1**. a) Sketch of the device consisting of a PMN-PT piezoelectric plate featuring pillars over which a WSe$_2$ ML is transferred. The top and bottom surfaces of the plate are gold-coated. The application of a voltage (electric field $F_p$ along the poling direction) through an AlN chip-carrier produces the out-of-plane ($\varepsilon_\perp$) and in-plane ($\varepsilon_\parallel$) deformation of the piezoelectric plate, which is eventually transferred to the attached ML. b) Optical-microscopy image of a WSe$_2$ ML (transparent brownish region delimited by a dashed white line) transferred on top of the piezoelectric pillars. c) 3D AFM picture of a representative piezoelectric pillar (height ~ 115 nm and diameter ~ 1400 nm) covered by a WSe$_2$ ML. d) Micro-PL map of a WSe$_2$ ML transferred on top of piezoelectric pillars taken on the region highlighted by the point-dashed black box in panel b). The maximum intensity of the PL spectrum is reported for each pixel and normalised to the absolute maximum recorded over the entire map.

Dashed white circles highlight the position of the four pillars. e) Micro-PL spectra recorded outside (orange line) and at the edge (dark-cyan line) of a pillar. The emission energy of the WSe$_2$ ML 2D-neutral-exciton is indicated with a blue vertical line. f) Second order autocorrelation measurement obtained on a single QE. The single photon emission nature is confirmed by the low value $g^{(2)}(0) = 0.21(3)$. The orange solid line shows the fit to the experimental data.[35]

Figure 1b shows an optical microscopy picture of the device for a pitch distance between pillars of 5 μm. We highlight that, besides inducing the formation of QEs in the ML, the piezoelectric pillars can deform themselves to dynamically reconfigure the strain state of the attached ML. Figure 1c shows the AFM image from one of the pillars covered with the ML, which results in being firmly attached to the surface of the piezoelectric plate. Interestingly for what will be discussed in the following, nanometre-sized protuberances (height ~10 nm and diameter ~ 50 - 100 nm) are formed along a corona over to the pillars' edges. This is ascribed to ML bending at imperfections formed during the device fabrication and are likely to be responsible for the formation of QEs, as explained below.

To investigate the formation of site-controlled QEs, we performed spatially resolved micro-photoluminescence (PL) measurements at $T = 5$ K (see methods). Figure 1d shows the PL intensity map as obtained by detecting light in the spectral window 710-780 nm and taking the maximum value of the PL spectrum while scanning the excitation laser spot in steps of about 110 nm around four pillars. Local PL intensity maxima attributed to the radiative recombination of localized excitons are found on top of the pillars, specifically around the corona. Despite these lines are related to exciton or multiexciton complexes (such as trions, biexcitons, etc..) we generally refer to them as localized excitons. We also note that the contribution of the 2D neutral free exciton (expected at around 1.74 eV, see Figure 1e) is negligible for the excitation power used in the experiments. Representative micro-PL spectra acquired outside and at the edge of a pillar are shown in Figure 1e. In the latter, a relatively sharp (full width at half maximum of ~ 370 μeV) emission line is observed in the spectral range where QEs in WSe$_2$ usually appear.[3,5] To demonstrate that these isolated lines act as single photon sources, we spectrally filter one of them and perform second-order photon-correlation measurements $g^{(2)}(\tau)$ using a standard Hanbury-Brown and Twiss set-up. As shown in Figure 1f, pronounced photon antibunching is observed, with values at zero-time delay as low as $g^{(2)}(0) = 0.21 \pm 0.03$. We repeated the experiments for several emitters and observed similar results, thus unequivocally demonstrating that these recombination centres act as non-classical light sources. We attribute the origin of these QEs to the static strain gradients that drive the photo-generated excitons into potential wells forming at the border of the top part of the pillars, specifically at the protuberances visible in the AFM image of Figure 1c.[31] It is important to mention that our results agree with previously reported studies that use pillars of similar size and aspect ratio.[8] However, here the pillars are made from a piezoelectric material that can deform its shape and thus enable dynamic studies of the optical properties of the QEs.

To investigate the type of strain introduced by the piezoelectric pillar, we first perform finite-element-method (FEM) numerical simulations of a PMN-PT plate featuring pillars on top (no ML attached) as a function of the electric field $F_p$ applied along the poling direction. We stress that the protuberances at the pillars' rim (see Figure. 1c) are not considered in these simulations, but they will be the focus of the next section. **Figure 2a** shows the hydrostatic in-plane strain $\varepsilon_{xx} + \varepsilon_{yy}$ map for $F_p = 15$ kV cm$^{-1}$. A non-uniform strain distribution is observed on top of the pillar, with compressive strain values varying from about -0.25% at the centre to about -0.07% towards the edges, where strain-relaxation is at play. The out-of-plane strain $\varepsilon_{zz}$ is also slightly compressive, as shown in Figure 2b together with the morphological changes of the pillar upon the application of different $F_p$. Taking into account that a compressive/tensile strain introduces an increase/decrease of the WSe$_2$ band-gap, we expect a blueshift of the QE emission lines when we apply positive voltages, with values that depend on their specific location across the pillar (see Figure 2c for the points highlighted with letters in Figure 2a). This is exactly what we observe experimentally, as shown in Figure 2d. We measure shift rates up to 0.3 meV cm kV$^{-1}$, with differences of up to one order of magnitude for distinct QEs (see the inset of Figure 2d). Before proceeding further, we would like to emphasize the relevance of this result: First, in stark contrast to previous works focusing on naturally occurring wrinkles onto piezoelectric actuators (without pillars),[15] we always observe that the energy shift is following the sign of the $F_p$. This is readily explained by the simulations shown in Figure 2a that, in combination with Figure 2c,d, also suggests that the ML closely follows the pillar shape. Second, we highlight that the energy shift is fully reversible (see Supporting Information (SI), Note 1) and, when we use both positive and negative $F_p$, we can observe energy variations up to 8 meV (see Figure S2.3b,h). These features of the device are the key to obtain the following main result of our work.

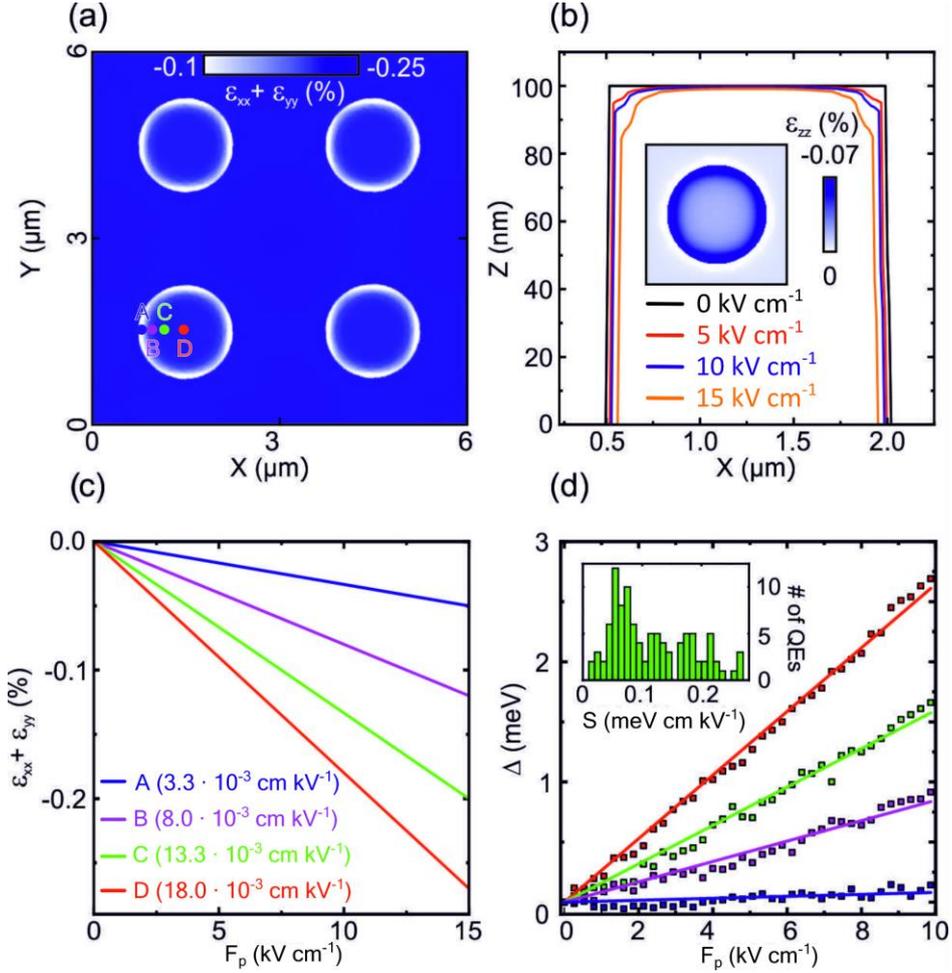

**Figure 2.** a) Numerical FEM simulation of the in-plane hydrostatic strain $\varepsilon_{xx} + \varepsilon_{yy}$ as obtained in a piezoelectric (001)-PMN-PT plate with pillars and for an applied electric field $F_p = 15$ kV cm$^{-1}$. b) Profile of the pillar along the Z-X directions and as a function of the electric field applied to the piezoelectric plate. The insert shows the FEM simulation of the out-of-plane strain $\varepsilon_{zz}$ for $F_p=15$ kV cm$^{-1}$. c) In-plane hydrostatic strain $\varepsilon_{xx} + \varepsilon_{yy}$ taken from numerical simulations as a function of $F_p$. The different lines correspond to the strain measured at the different points highlighted with letters in a). d) Shift of the energy of QEs located at different positions on top of the piezoelectric pillar as a function of $F_p$. The different coloured points correspond to different QEs. The solid lines are linear fits to the experimental data. The insert shows the histogram of the energy-shift rate of all the QEs investigated in this work.

## 2.1. Strain-induced redistribution of excitons among localized quantum emitters.

We now investigate the evolution of the PL intensity with $F_p$ to understand whether strain can be used to control the population of QEs in TMDs. **Figure 3a** illustrates a series of spectra collected while varying $F_p$ in the range 0 – 15 - 0 kV cm$^{-1}$. Two emitters, labelled as QE1 and QE2, exhibit an interesting behaviour: while $F_p$ increases, the maximum PL intensity of QE1 (QE2) decreases (increases). This feature is clearly visible in Figure 3b, which depicts the two spectra at 0 and 15 kV cm$^{-1}$. The intensity of QE1 drops by a factor 0.6 of its initial value at 15 kV cm$^{-1}$, while QE2 increases by a factor 3. Additionally, these intensity modulations are reversible. Figure 3c,d shows the maximum PL intensity in the spectral range of QE1 and QE2, respectively, as a function of $F_p$. The black and red dots, which correspond to the forward and backward ramps, overlap, suggesting the reversibility of the QEs' brightening/darkening.

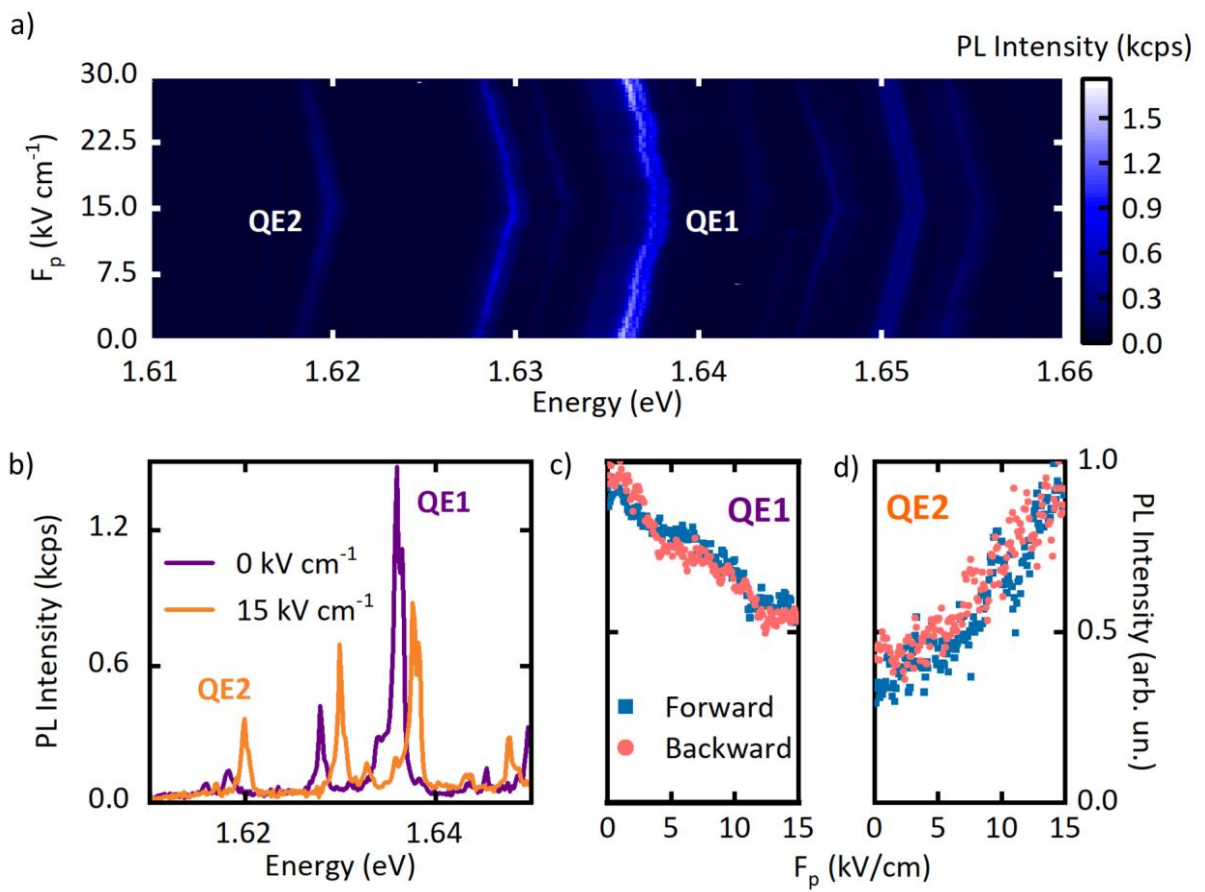

**Figure 3.** a) Colour-map of the maximum PL intensity as a function of the external electric field $F_p$. The spectra are collected on a pillar while a back-and-forth scan of $F_p$ from 0 to 15 kV cm$^{-1}$ is performed. b) PL spectra at 0 and 15 kV cm$^{-1}$. A QE1 (QE2) darkening (brightening) is observed at 15 kV cm$^{-1}$. c) Maximum PL intensity as a function of $F_p$ for QE1 and d) QE2, respectively. The intensity of the two emitters shows two main features: (i) while QE1 decreases for increasing $F_p$, QE2 increases, and (ii) the intensity of both QEs goes back to the initial value, suggesting that the process is reversible.

To substantiate these findings, we systematically investigate the evolution of the PL intensity with $F_p$ over one specific pillar, - pillar 4 in Figure 1d - (see **Figure 4a**), but very similar results were observed in the other 3 pillars (see SI, Note 2). Figure 4e shows five PL spectra collected from -10 to 30 kV cm$^{-1}$ in steps of 10 kV cm$^{-1}$ collected from the white spot in Figure 4a. Stark changes are observed. Besides the systematic energy shift discussed in the previous section, we find that QEs strongly change their intensity, some even disappearing. These changes are particularly apparent for the two emission lines indicated by QE2 and QE3 in Figure 4e. As we are dealing with intensity measurements, particular care must be devoted to the alignment and stability of the collection optics. We must exclude, for example, drifts of the sample with respect to the objective collecting light while the pillar is expanding/contracting. We adopt several preventive measures to verify this is not the case in our experiments. First, we utilize an experimental set-up in which the objective is held at a low temperature and in a vacuum inside the cryostat, together with the sample. This ensures long-term stability, and we can collect light from the same QE for days without the need to re-adjust the sample position. Second, we record several PL maps (like the one shown in Figure 4a) for each applied $F_p$ and we find that the panchromatic integrated intensity – as obtained by integrating the entire PL spectrum over all the pixels inside the region of interest (Zone 1 and Zone 2, determined as the dashed green and purple ellipses in Figure 4a) — remains almost constant (see data points in Figure 4f, where the standard deviation over all the data is below 5 %). This clearly indicates that the number of detected photogenerated excitons does not vary with $F_p$ and that no misalignment of the collection optics is affecting our experiments. Third, we performed a careful analysis to demonstrate that blinking at a short time scale does not affect the estimation of the QE intensities (see SI, Note 3), which we can provide with about 15% uncertainty. Finally, to exclude unequivocally that the intensity changes are related to sample drifts, we perform multi-Gaussian fits of the PL spectrum recorded in each position of the spatial map. Then we record the intensity of the light emitted by the single QE, namely the intensity of the Gaussian peak of a given energy, in each position over the whole 2D map and for each specific $F_p$. This procedure, which follows the one reported in Ref. [36] and we called as hyperspectral maps, leaves us with new maps in which we can single out the emission area of each specific QE, which corresponds to the area of the device where each specific QE generates photons. We then combine the individual maps of each QE for each $F_p$, as shown in Figure 4b,c,d for QE1 (1.644 eV at $F_p = 0$), QE2 (1.691 eV at $F_p = 0$), and QE3 (1.672 eV at $F_p = 0$) acquired at $F_p = -10$, 10, and 30 kV cm$^{-1}$. Each map of individual QEs is separately normalized to its maximum along the electric field sweep. We note that the size of these areas is determined by the exciton diffusion length and the diffraction limit, about 300 nm for the optics and laser used in this work, see methods. We emphasize that, although we cannot determine the exact position of the emitters due to their proximity, it is possible to distinguish them by their optical emission energy in the PL spectra. For further considerations about the strain effects on the QEs' position, see SI, Note 4. With these maps at hand, we can extract both the total integrated intensity from a single QE (Figure 4g), as well as the energy shifts of each QE (Figure 4f) for each value of $F_p$. It is evident that QE2 (QE3) dramatically increases (decreases) in intensity with $F_p$, while QE1 remains almost constant. Similar intensity changes are observed for other QEs visible in the spectra shown in Figure 4e (note that it is not always possible to follow their complete evolution with $F_p$, as some lines merge with others). Yet, these measurements show that piezoelectric-induced strain can modify the intensity of each emission line dynamically. Remarkably, QE3 emission is completely switched off for applied voltages above 20 kV cm$^{-1}$. It is also worth noticing that no clear correlation between the energy shift and the rate at which the PL intensity changes are observed. Besides lines that change intensity in a non-linear fashion but at a constant energy-shift rate, we also found QEs with similar energy-shift rates (like QE2 and QE3) but featuring opposite intensity changes.

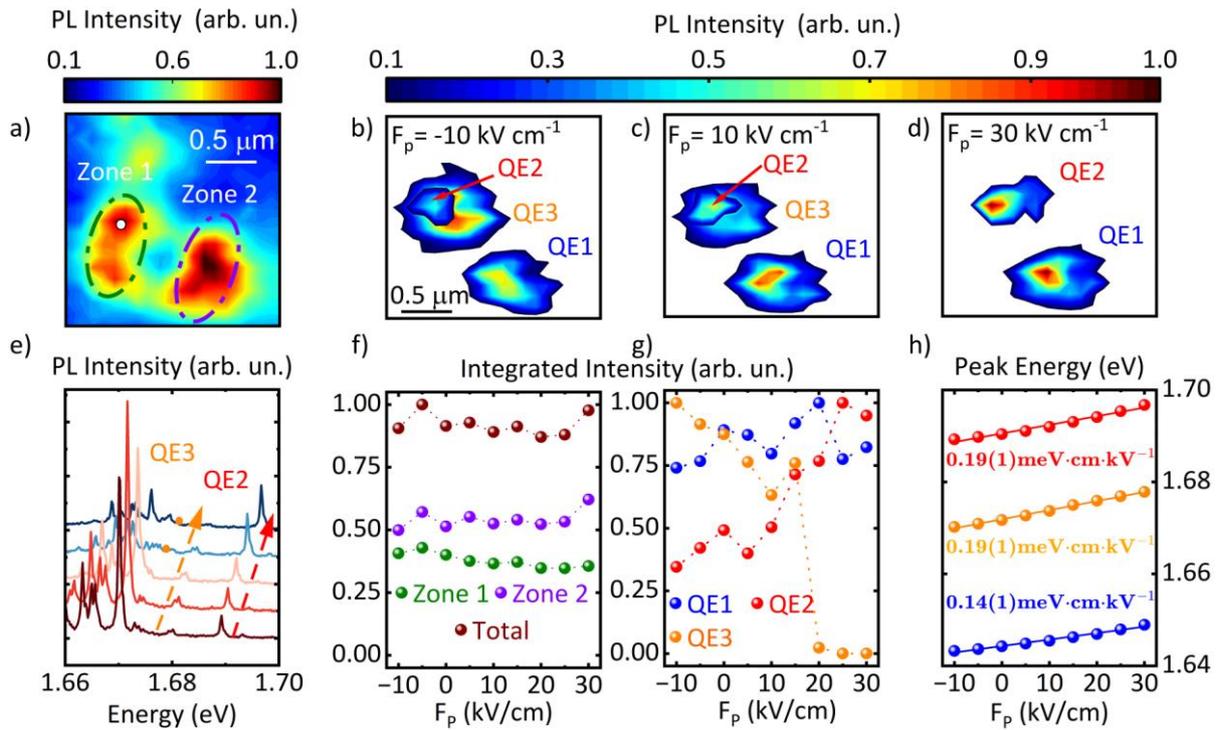

**Figure 4.** a) Spatially resolved micro-PL map of pillar #4 shown in Figure 1.d). b) Spatially resolved micro-PL map of light emitted by QE1, QE2 and QE3 for $F_p$ = -10 kV cm$^{-1}$. The maps are extracted separately from a hyperspectral map of PL spectra and making multi-Gaussian fits of the single PL spectrum in each pixel to identify the intensity of the light emitted by the single QE in each position of the whole 2D map. Each map of individual QEs is separately normalized to its maximum along the electric field sweep. The white area corresponds to regions where the QE's intensity is smaller than 10% of its maximum. c) The same as b) for $F_p$ = kV cm$^{-1}$. d) The same as b) for $F_p$ = 30 kV cm$^{-1}$. e) Micro-PL spectra recorded at the position marked by the white spot in a). The spectra show the evolution from -10 kV/cm to 30 kV cm$^{-1}$. The orange dots highlight the QE3 expected position, showing the emitter's disappearance. The dashed orange and red arrows guide eyes to follow QE2 and QE3. f) Integrated area of the panchromatic PL spectrum as a function of $F_p$ in different regions over the pillar. Brown dots refer to the total pillar area, while the green and purple dots refer to the two ellipses shown with the same colour in a), labelled as Zone 1 and Zone 2, respectively. g) Normalized integrated area of the PL intensity of the three QEs studied in b), c), d) as a function of $F_p$. The uncertainty in the intensities is estimated to be around 15%, as discussed

in SI, Note 3. h) Energy shifts of the three QEs studied in b), c), d) as a function of $F_p$. The solid lines are linear fits to the experimental data, while *m* stands for the energy shift rate.

At first sight, there can be different explanations for this phenomenon. For instance, one may argue that the band bending ensuing the application of $F_p$ can lead to a change of the exciton dipole moment and, possibly, a separation of electrons and holes that decreases the emission intensity and favours the appearance of charged exciton complexes.[37] Another possible explanation – in the well-established intervalley-defect-exciton picture[28] – is that strain modifies the coupling between weakly localized exciton states and strongly localized defect levels. Both effects should indeed lead to a change in the oscillator strength of the transitions. However, two important points need to be noticed here: First, the strain values delivered by our actuator are on the order of 0.1 - 0.3% and should lead to a relatively small change in the oscillator strength of the QE transitions[13,28] (see SI, Note 8 for further details). Second and most importantly, the overall integrated intensity of the light emitted by the QEs over all their extent remains constant. This is true not only for the whole pillar but also for spatial regions that have a size comparable to the exciton diffusion length, as shown in Figure 4f for the green and purple ellipses in Figure 4a. Thus, the fact that the QEs change intensity while the overall PL intensity from the region of interest remains constant strongly suggests that the induced strain is regulating the distribution of the photogenerated excitons among the different QEs without changing their actual number.

At this point, a rather simple and intuitive physical picture emerges: The QEs originate from localized potential wells (local band-gap minima) of a disordered potential landscape, similar to the case of random alloys.[38] Their populations, directly linked to their PL emission intensity, are related to the way excitons funnel and thermalize toward the potential-landscape minima, which are in turn determined by the local deformation of the ML right after the fabrication process, i.e., by *the static strain field*. Instead, the *dynamic strain* delivered by the piezoelectric pillar modifies the relative heights of different potential wells and the way photo-generated excitons distribute among the different QEs while keeping their overall number constant. The exciton distribution changes at each applied $F_p$, and the QE PL intensity modifies accordingly. To verify the correctness of this suggestive hypothesis, we perform theoretical calculations that combine drift-diffusion equations with the strain profile obtained by numerical FEM simulations, as described in the following.

We use the excitonic drift-diffusion equations considering a potential landscape $u(\vec{r})$ which changes as a function of the position $r$ across the pillar. For a non-uniform density of excitons $n(\vec{r})$, the equilibrium state condition for exciton diffusion $\vec{J_d} = D\nabla n(\vec{r})$ and drift current $\vec{J_\mu} = \mu n(\vec{r})\nabla u(\vec{r})$ is described by the following differential equation:[39-41]

$$\nabla(\vec{J_d}) + \nabla(\vec{J_\mu}) - \frac{n(\vec{r})}{\tau} - n^2(\vec{r})R_A + S(\vec{r}) = 0 \quad (1)$$

where $D$ is the diffusion coefficient, $\mu = \frac{D}{K_B T}$ is the mobility that depends on temperature, $\tau$ is the exciton lifetime, and $R_A$ is the Auger recombination rate that we assume to be negligible with respect to the radiative recombination rate due to the low power excitation used during the experiment[42] (see SI, Note 9 for further details). $S(r)$ is the exciton generation source, *i.e.*, our excitation laser, which, according to the experimental conditions, we assume to have a Gaussian profile $S(r) = \frac{I_0}{2\pi\sigma^2} e^{-\frac{r^2}{2\sigma^2}}$ with maximum intensity $I_0$ and $\sigma$ is the beam width radius. In Equation 1, the terms that dynamically drive the change in the exciton density are the diffusion and the drift terms. The generation and recombination terms are mainly responsible for exciton density in steady-state conditions, with the latter also partially responsible for the size of the potential landscape experienced by the exciton through diffusion length $l = \sqrt{D * \tau}$. On the other hand, the ratio between the diffusion coefficient and the mobility, which is basically related to the temperature, sets two different regimes: (*i*) the diffusive regime in which most of the excitons spread over the whole area - a condition that is usually achieved at elevated temperatures; (*ii*) at low temperature the drift regime rules the diffusion of excitons, which move towards the minima of the potential landscape. Since the experiments are performed at low temperatures (4 K), we can safely assume that the simulations are mainly concerned with the drift regime.

To calculate the exciton distribution $n(r)$, we first need to evaluate the potential landscape $u(r)$ upon the application of $F_p$, which is obtained by FEM simulations of the strain status of the piezoelectric pillar (see methods and SI, Note 7). More specifically, we first calculate the hydrostatic strain distribution map ($\varepsilon_{xx} + \varepsilon_{yy}$) and then obtain the bandgap energy variation with respect to the one corresponding to the unstrained ML via (see SI, Note 6):

$$E_g(eV) \simeq E_{go} - 0.0638\varepsilon_{xx} - 0.0636\varepsilon_{yy} \quad (2)$$

where $E_{go}$ is the "unstrained" bandgap energy.[43,44] Note that the contribution of $\varepsilon_{zz}$ is found to be negligible and is therefore neglected in the calculation.

In this step of our theoretical approach, there are a few points that need to be discussed: First, we now consider the geometrical protuberances at the pillar rim that simulate the disordered profile (visible in the AFM map of Figure 1c) leading to the formation of the different QEs. Second, we highlight that it is extremely difficult to experimentally evaluate the initial *static* strain configuration (or pre-strain) at the QE location with the desired accuracy, as it results right after mechanical exfoliation. This would require optical spectroscopy techniques with sub-diffraction limited resolution at low temperature, compatible with the application of high voltages, and with sufficient stability and throughput to enable intensity measurements with high accuracy. Thus, we can evaluate only the *dynamic* changes of $u(r)$ induced by $F_p$, i.e., by the strain fields induced by the piezoelectric actuator. This is not a problem *per se*, but our model requires the initial height of the potential well leading to the QE formation as input. Considering that the QEs are formed at the geometrical protuberances along the periphery of the pillar (see Figure 1c) and that the emission energy is approximately equal to the local bandgap value of the ML at these positions, the height of the potential wells can be indirectly estimated via Equation 2 and assuming the induced strains by the protuberances $\varepsilon_{tot} = \varepsilon_{yy} + \varepsilon_{xx}$. More specifically, using the spread in emission energy of the different QEs (which varies in a range from 0.04 - 0.1 eV with respect to the unstrained ML emission energy of the 2D-exciton, about 1.75 eV) Equation 2 shows that the pre-strain in the ML at the QEs positions ranges from about 0.3 % to 0.7 % (we did not consider quantum confinement effects in our model, as the withdrawn conclusions are qualitatively similar). These pre-strain values are thus randomly assigned to the different protuberances, whose geometries have been included in the FEM simulations according to the results of AFM measurements (see SI, Note 7 for more details). Since it is reasonable that the ML closely follows the piezoelectric substrate, we assume it feels the same strain as $F_p$ varies. Finally, we note that the pillar structure is not pivotal in redistributing excitons between QEs. Indeed, the pillar initially deforms the ML leading to a bandgap gradient in the ML from the bottom (higher) to top (lower) (see SI, Note 8), and thus a net excitons funnelling towards the top of the pillar where they recombine at QEs positions (see Figure 1d). We therefore limit our study to the diffusion of excitons between QEs placed along a circle mimicking the geometrical distribution of the protuberances on the pillar rim.

With *u(r)* at hand, we can now solve the drift-diffusion Equation 1 and map the exciton distribution *n(r)* across the whole potential landscape, a procedure repeated at $F_p$ = 50 kV cm$^{-1}$ and 25 kV cm$^{-1}$. Left half of **Figure 5a** shows the results for *n(r)* at $F_p$ = 50 kV cm$^{-1}$. To reproduce our experimental settings, we performed a convolution between *n(r)* and a point spread function (PSF) with 250 nm FWHM to match the typical spatial extent of our QEs (as measured after spectrally isolating the different emission lines, see Figure 4b,d and S2.1-3). Right half of Figure 5a shows the results of the convolution. The PSF enlarges the peaks in *n(r)* (corresponding to the protuberances), resulting in a profile which is similar to what was observed experimentally (as reported in Figure 1d, for example). Figure 5b depicts the difference between the exciton densities *n(r)* at $F_p$ = 50 and 25 kV cm$^{-1}$ after the PSF convolution.

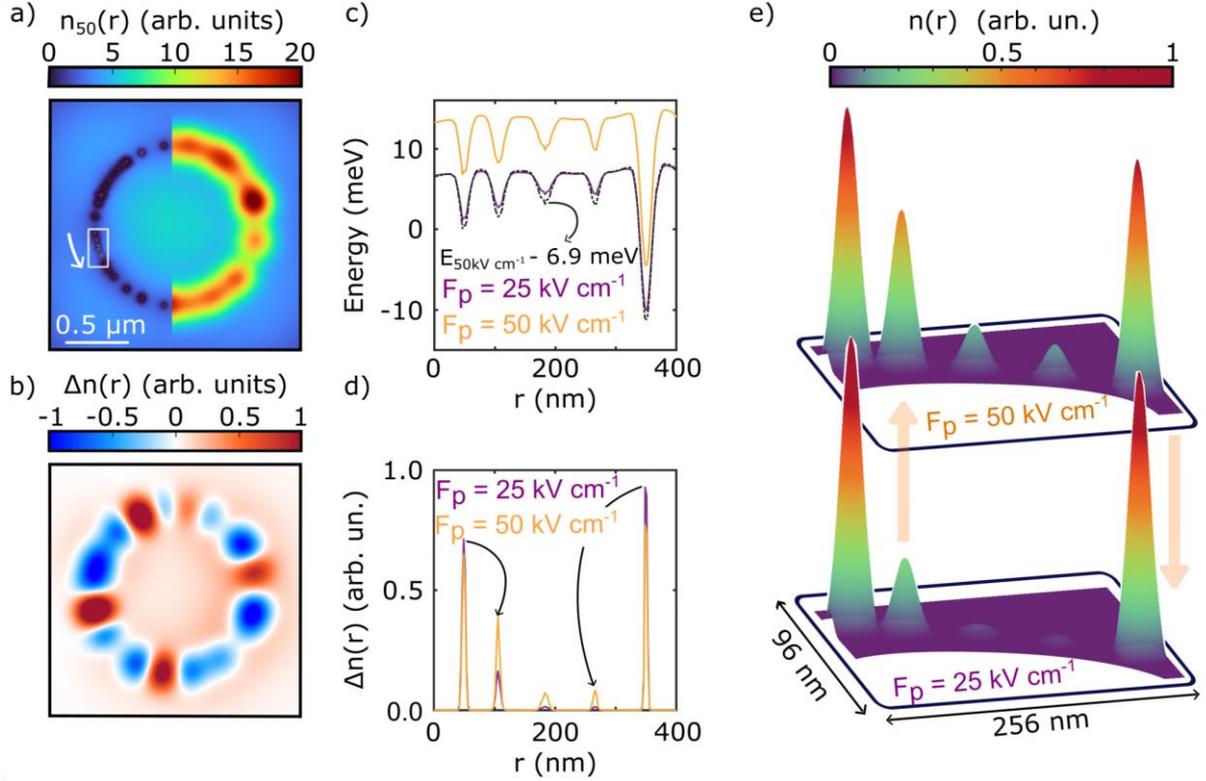

**Figure 5.** a) Exciton density distribution at $Fp$ = 50 kV cm$^{-1}$ with and without convolution with the point-spread function (PSF). The 250 nm PSF reproduces the effect of the collection optics. The result without convolution is the left portion of the image. The white arrow indicates the arc of circumference along which band gap variation and exciton density (without convolution) are extracted to build up panel c). The white rectangle highlights the region selected for panel d), where both defects gaining and losing excitons occur. b) Variation of the exciton density distribution, obtained by subtracting the exciton density at $F_p$ = 25 and 50 kV cm$^{-1}$ after the PSF convolution. c) Band-gap energy variation and d) exciton density as recorded along the white arrow in a) for $F_p$ = 50 kV cm$^{-1}$ and $F_p$ = 25 kV cm$^{-1}$. For clarity, the black dashed line shows the energy shift obtained for $F_p$ = 50 kV cm$^{-1}$ red-shifted by 6.9 meV. e) Schematics of the 3D exciton density for the five protuberances highlighted by the white box in panel a) at $F_p$ = 25 and 50 kV cm$^{-1}$ after a gaussian smoothing. The three central protuberances gain excitons while the two later ones lose carriers.

We observe stark changes: some protuberances collect more/fewer excitons at the expense/in favour of others while the overall number of excitons across the profile remains constant - in perfect qualitative agreement with our experimental results. To obtain further insight into this effect, we show in Figure 5c,d the values of the band gap variation and *n(r)*, without PSF convolution, calculated along a circular corona parallel to the white arrow reported in Figure 5a for both the values of $F_p$. It is apparent that the potential wells whose height/depth increases more with compressive strain gain excitons at the expense of nearby wells, as indicated by the black arrows in Figure 5d. However, we emphasise that the relative height of the potential wells is not the only relevant parameter, but also the relative distance among the different protuberances plays a role in the exciton diffusion. Moreover, focusing on just a few protuberances (see Figure 5e), we can clearly observe cases in which each of them gains/loses excitons at a different rate, a behaviour that we also observe experimentally (see SI, Note 2 and 5).

**Conclusions**

In summary, our results show that the strain fields delivered by piezoelectric pillars allow the energy and the intensity of light emitted by QEs in TMDs MLs to be actively controlled. The experimental findings are explained by a strain-induced modification of a disordered potential landscape, which enables a redistribution of the excitons among different QEs. It is worth emphasizing that we can explain the experimental findings without invoking a strain-induced modification of the coupling between the monolayer exciton states and the localized defect states. In fact, the change in the oscillator strength of the optical transitions is supposed to be small for the strain values (about 0.1 - 0.3%) investigated in this work. Yet, devices that enable full control over the strain tensor in TMDs with magnitude as large as 1% are available,[45] and we envision the development of integrated quantum photonic devices to control the flow of excitons among distinct and interconnected QEs in TMDs, and eventually, to switch them on and off on demand. Moreover, our device concept could be used to improve the brightness of single photons sources in 2D materials and opens the path towards their exploitation in quantum communication applications.[43]

**Methods**

*Sample fabrication:* The piezoelectric pillars were fabricated on top of a 200-m-thick PMN-PT piezoelectric plate by a focused ion beam (FIB). The patterning of the piezoelectric substrates was performed using a cross-beam FIB/SEM microscope. The Ga+ Ion column was used for this purpose under an acceleration voltage of 30 kV with a current probe of 84 pA. Binary image-based masks for the definition of the circular pillar arrays have been created with an image pixel size of 5 nm in total areas of 18·18 µm². Pillars with a diameter of 1400 nm were patterned. The milling rate was previously calibrated by creating cross-sectional cuts in the sample leading to total exposure times per 18·18 µm$^2$ areas of 533 seconds for steps around 100 nm. Afterwards, both sides of the patterned piezoelectric plate were coated by electron-beam and thermal evaporation with a Cr(5 nm)/Au(40 nm) bilayer for electrical contact. The WSe$_2$ ML was obtained by mechanical exfoliation on PDMS stamps and transferred on top of the gold-coated piezoelectric nanopillars by using the dry transfer technique.[47] A positive voltage was used for the poling of the piezoelectric plate, meaning that positive/negative voltages introduce an in-plane compressive/tensile strain field on its surface. A sketch of a complete device is shown in Figure 1a. In particular, the strain fields are introduced on the attached ML by applying voltages on the bottom side of the piezoelectric plate while the top side is grounded.

*Photoluminescence:* A continuous-wave He-Ne laser emitting at 633 nm is used as excitation laser source for the µ-PL experiment. The laser passes through a beam-splitter, which reflects 10 % of laser light, and then it is sent through a periscope inside a He close-cycle cryostat equipped with an objective with 0.81 numerical aperture. The sample is mounted on an x-y-z piezoelectric stage and is electrically connected to an outer voltage supplier via high voltage vacuum-feedthroughs. The PL emitted by the sample is then collected by the same objective in a back-scattering geometry, and the collimated beam is spatially filtered via a single-mode optical fiber. A long pass filter filters out the laser at the spectrometer entrance, which is 750 mm long and equipped with a 300 g mm$^{-1}$ diffraction grating. At the outport of the spectrometer, a liquid nitrogen-cooled CCD acquires the PL signals. The sample is mounted underneath the objective (kept fixed) and moved with nanopositioners in steps of about 110 nm. In this way, a PL spectrum is taken from every single pixel of the 2D map. The time correlation measurements are performed using a Hanbury-Brown and Twiss setup consisting of a beam splitter and two avalanche photodiodes (time resolution of about 400 ps) connected to correlation electronics.

*Numerical simulations by finite element method (FEM):* The in-plane hydrostatic strain field ($\varepsilon_{xx}+\varepsilon_{yy}$) distribution on PMN-PT piezoelectric 5x5 mm$^2$ plates - (001) orientation - is simulated by FEM using the commercially available software COMSOL Multiphysics. A 3D pillar model is built with a geometry inspired by AFM measurements. We placed different ellipsoids to model the nanometer-sized protuberances (which lead to the formation of QEs) along the pillar rim. The voltage is applied on the bottom side of the piezoelectric plate, inducing an out-of-plane electric field across the thickness of the piezoelectric plate, [001] direction, which is varied from 25 to 50 kV cm$^{-1}$. The top side of the piezoelectric plate is set to ground. The gluing effect of the device to a chip carrier (obtained using silver paint in the experiments) is simulated by including an isotropic solid material ("soft block") under the piezoelectric substrate with small Young modulus (about 500 Pa). The latter value has been checked in a separate work by direct comparison with experiments where the PL shift versus measured strain values have been analyzed.[47] The bottom boundary of the "soft block" is fixed. See SI, Note 7 for further information about the geometrical design and specific values for the elastic and piezoelectric tensor used for the PMN-PT plate.

*Drift-diffusion model (DDE):* The drift-diffusion Equation 1 is solved using the commercially available software COMSOL Multiphysics through the coefficient form PDE module. The details of the model can be found elsewhere.[39-41] We set the exciton density equal to zero at the borders, defined by a circle centred on the pillar's centre and radius equal to four times the pillar's radius (2.8 μm). Our boundary conditions are realistic as, on the borders, the distance from potential minima is much greater than the diffusion length of the carriers (200 - 300 nm). To follow the experimental procedure, we illuminate a point of the potential landscape with the Gaussian laser source *S(r)*, and we then run the drift-diffusion simulation to calculate *n(r)*. We iterate the procedure scanning the laser source origin in steps of 200 nm over a square grid with an edge equal to 2 μm, and we then sum all the obtained *n(r)*, convolving their weights with a gaussian point spread function with 250 nm FWHM. The parameters used in the simulations are the one reported in the following. $D = 0.6$ cm$^2$ s$^{-1}$, $\tau = 1.1$ ns, $T = 10$ K, $FWHM_{Laser} = 2 * \sqrt{2 \ln \ln (2)} * \sigma = 0.5$ μm. Reported values of diffusion coefficients for excitons in monolayer TMDC range from $D = 0.1$ cm$^2$ s$^{-1}$ to $D = 10$ cm$^2$ s$^{-1}$.[48] For the lifetime value we chose lifetime of the same order of magnitude we measured in QEs in our sample, see SI, Note 9. It is worth mentioning that we run simulation with different lifetime values, spanning from ten ps to ten ns. Since lifetime rules mainly the number of excitons, and partially the potential landscape felt by the excitons, only changes in the exciton density absolute numbers were observed.

*Energy band-gap calculation in strained WSe$_2$ monolayers:* To calculate the dependence of the bandgap on the strain fields introduced by the piezoelectric device, we performed ab-initio simulations on strained WSe$_2$ monolayers with the Quantum ESPRESSO package.[49] We used ultrasoft pseudopotentials[50] and the GGA approximation with the PBE parametrization[51] to treat exchange and correlation. A plane-wave energy cutoff of 55 Ry and a 7x7 k-point grid on the *xy* plane are used in the calculations. The coordinates were relaxed until all forces were smaller than 10$^{-3}$ Ry Bohr$^{-1}$. The resulting band structure without strain has a direct bandgap of 1.53 eV. Notice that although the absolute value of the bandgap given by GGA is usually underestimated, the relative evolution with strain can be adequately captured.[49] Consequently, the actual strain field value can be calculated by introducing an offset. Next, we calculated the evolution of the bandgap change ($\Delta E$) as a function of uniaxial $\varepsilon_{xx}$ and $\varepsilon_{yy}$ components of the strain fields (note that the $\varepsilon_{zz}$ component is found to be negligible). Such evolutions are roughly linear, and by interpolating them, it is possible to obtain the following expression ($\Delta E$ in eV and strains in %): $\Delta E(\varepsilon_{xx}, \varepsilon_{yy}) = -0.0638364 \cdot \varepsilon_{xx} - 0.0636273 \cdot \varepsilon_{yy}$, where we have taken as the coordinate at the origin the average of the coordinates at the origin of both uniaxial interpolations. Using this formula, it is possible to reproduce the band-gap differences for arbitrary strain values with reasonable precision. The graph of the interpolation can be found in the SI, Note 6.

**Supporting Information**

Supporting Information is available from the Wiley Online Library or from the author.

**Conflict of Interest Statement**

The authors declare no conflict of interest.

**Data Availability Statement**

The data that support the findings of this study are available from the corresponding author upon reasonable request.

**Acknowledgements**

G. R., A. M. S., and D. T. contributed equally to this work. We acknowledge Javier Taboada-Gutiérrez for his helpful contribution at the initial stage of this work. This work was financially supported by the MUR (Ministero dell' Università e della ricerca) via the FARE project 2018 n. R188ECR2MX (STRAIN2D), and the QuantERA project, QUANTERA Call 2021, EQUAISE. J.M.-S. acknowledges financial support from the Ramón y Cajal Program of the Government of Spain and FSE (RYC2018-026196-I), the Spanish Ministry of Science and Innovation (State Plan for Scientific and Technical Research and Innovation grant number PID2019-110308GA-I00/AEI/10.13039/501100011033) and project PCI2022-132953 funded by MCIN/AEI/10.13039/501100011033 and the EU "NextGenerationEU"/PRTR". A.H.-R. and S.M. acknowledge support from the European Union's Horizon 2020 research and innovation programme under Marie Skłodowska-Curie grant H2020-MSCA-IF-2016-746958.

A.H.-R. acknowledges funding from the Spanish AEI under project PID2019–104604RB/AEI/10.13039/501100011033.

Received: ((will be filled in by the editorial staff))
Revised: ((will be filled in by the editorial staff))
Published online: ((will be filled in by the editorial staff))